\documentclass{article} 
\usepackage{nips15submit_e,times}
\usepackage{hyperref}
\usepackage{amsmath}
\usepackage{booktabs}
\usepackage{url}
\usepackage{color}
\usepackage{footnote}
\makesavenoteenv{tabular}
\makesavenoteenv{table}

\newcommand{\rpm}{\raisebox{.2ex}{$\scriptstyle\pm$}}

\title{What do Vegans do in their Spare Time?\\
Latent Interest Detection in Multi-Community Networks
}

\author{
Jack Hessel, Alexandra Schofield, Lillian Lee, David Mimno \\
Cornell University\\
\texttt{\{jhessel, xanda, llee\}@cs.cornell.edu, mimno@cornell.edu} \\
}

\nipsfinalcopy 

\begin{document}

\maketitle

\begin{abstract}

Most social network analysis works at the level of interactions
between users. But the vast growth in size and complexity of social
networks enables us to examine interactions at larger scale. In this
work we use a dataset of 76M submissions to the social network Reddit,
which is organized into distinct sub-communities called subreddits.
We measure the similarity between entire subreddits both in terms of
user similarity and topical similarity. Our goal is to find community
pairs with similar userbases, but dissimilar content; we refer to this
type of relationship as a ``latent interest.'' Detection of latent
interests not only provides a perspective on individual users as they
shift between roles (student, sports fan, political activist) but also
gives insight into the dynamics of Reddit as a whole. Latent interest
detection also has potential applications for recommendation systems
and for researchers examining community evolution.
\end{abstract}

\section{Introduction and Related Work}

In many social networks consisting of people and the relationships between
those people, there is additional {\em group} or {\em community}
structure.
This structure may be implicit and need to be inferred
\cite
{newman2006modularity, fortunato2010community}, or explicit, such as in
social-media sites offering user-created interest groups.
We focus on detecting \emph{anomalous} relationships between entire explicit
communities. In particular, we define relationships between
communities in two different ways, and use these competing definitions
to explore a phenomenon we refer to as ``latent interest.'' First, we
consider a simple measure based on user overlap: two communities are
similar if they
have many participants in common;
This measure has appeared in other work, such as \cite
{Hohwald2009intercommunity,olson2015navigating}.
Second, we define a
measure based on the language of each community.
By
explicitly comparing the differences between the user-based and
language-based metrics,
we can discover relationships that might not be captured by
using only a single similarity metric. For example, we can ask ``what
do vegans do when they aren't talking about veganism?''

Why might someone care about latent interests? Identifying a community
with a latent interest in another could assist in suggesting
interesting new communities for a user to join. Previous
recommendation systems are based on learned user similarities
\cite{xiao2001measuring, ziegler2004analyzing} or learned item
similarities \cite{sarwar2001item, linden2003amazon}, but they are
generally based on only one measure of similarity
(e.g., both users
watched a movie, both users purchased an item).
In our case, while such suggestions could be made solely based on
either the user similarities \emph{or} the content similarities,
identifying latent interests can produce recommendations that
incorporate \emph{both} while allowing flexibility in the trade-off
between novelty and similarity. Furthermore, detecting subtle
relationships between sub-communities might be useful for learning
about the context surrounding the evolution of single communities
\cite{palla2007quantifying}, or the adoption and abandonment of a
community by groups of users \cite{danescu2013no};
or it could be used to determine whether two sub-communities could be merged
\cite{Hohwald2009intercommunity} or to help infer whether there are
unobserved
ties between individuals in the two distinct sub-communities.

While we are aware of no work that \emph{contrasts} multiple
definitions of similarity to detect new types of relationships,
previous work has explored the interplay of topic and social
structure. For instance, in \cite{romero2013interplay}, the authors
examine the capacity of Twitter hashtags to predict underlying social
structures. In a similar vein, \cite{Kwak:ProceedingsOfWww:2010}
explore the extent to which Twitter acts as a news source and,
separately, as a social network. Also,
\cite{Ferrara:2014:OPT:2631775.2631808} explore clustering through
content and social structures using social tags on Instagram.

There also exist several topic models for uncovering latent network
structure that take content and social structure into account. For
instance, Topic-link LDA \cite{liu2009topic}, Pairwise Link-LDA
\cite{Nallapati:2008:JLT:1401890.1401957}, and Relational Topic Models
\cite{chang2009relational} jointly model social structures and user
content. Reddit has been specifically examined recently using backbone
networks \cite{olson2015navigating} but without looking deeply at
textual content.

We find that our methods for defining user and content similarity are
meaningful in a prediction setting, and then derive a heuristic method
for combining our measures to detect latent interests.

\section{Dataset Description}

We use a dataset of posts from \texttt{reddit.com} compiled by Tan and
Lee \cite{tan+lee:15} from an original data dump by Jason Baumgartner.
This dataset consists of roughly 76M submissions made to the social
networking website from January 2008 to February 2014. Notably, this
dataset does not include the user comments in response to these
submissions.
  Items by bots and spammers have been filtered out.
Reddit
is organized into a large number of interest-specific
subcommunities
called subreddits. A user may post to individual subreddits and
participate in the community upvoting, downvoting, and commenting on
content other users have submitted. An example of a popular subreddit
is \texttt{/r/aww}, where users submit pictures of cute animals.

For our analysis, we focus on subreddits that have enough text data to
understand the language used by members of the community. Hence, from
the set of all subreddits, we select subreddits for which there are
at least 500 text posts available, and at least 300 unique users have
submitted either text or links. This filter results in our final set
of 3.2K considered communities.

Next, we extract all 22.8M text posts made to our subreddit set.  Some
communities are much larger than others: the subreddit
\texttt{/r/leagueoflegends} contains more text posts than the smallest
800 communities we consider combined. To prevent our language model
from being overwhelmed by these large communities, we impose an upper
bound on the total number of text posts we model for a single
subreddit. Specifically, if a subreddit is associated with more than
5000 text posts, we select a random subset of 5000 of its posts to
consider. As a final filter, a text post is only considered if it has
a length greater than 20 words. After this filtration process, we are
left with just under 6.6M text posts.

We are interested in determining a group of users who have
participated in each community. For each user who has posted something
to any of our 3.2K subreddits, we extract the sequence of subreddits
they post to, as in \cite{tan+lee:15}.  For the purposes of this work
we discard posting order and frequency.

To encourage other researchers to consider networks of communities,
bigger and better corpora for topic modeling, and the interplay
between content and users, we publicly
release\footnote{\texttt{http://www.cs.cornell.edu/{\raise.17ex\hbox{$\scriptstyle\mathtt{\sim}$}}jhessel/projectPages/latentInterest.htm}}
the data. Specifically, we
release a version of the balanced, 6.6M
document corpus from \cite{tan+lee:15}, our hand-curated set of
overlapping subreddit communities, and the pairwise topic/user
similarities we used to define our networks.

\section{From Data to Graphs}
\subsection{Content Similarity}

We use topic models to define the content distances between all pairs
of subreddits
in our data.
Topic models are unsupervised matrix factorization
methods which assume hierarchical latent structure to data. Though
these models can be applied to many types of discrete input, they were
born out of a desire to understand topical themes within large textual
corpora. When applied to text, the most popular topic model,
 Latent
Dirichlet Allocation (LDA) \cite{blei2003latent}, assumes a set of
latent ``topics,'' represented by multinomial distributions over
words.
 Documents are
similarly represented as multinomal distributions over topics.  In
total, each word in each document is asssumed to be generated by first
drawing a topic from the document-level topic distribution, and then
drawing the specific word-type from a corpus-level topic-word
distribution. By adding a Dirichlet prior to these multinomial
distributions, LDA extends simpler models like probabilistic latent
semantic indexing \cite{hofmann1999probabilistic} to a fully
generative model, allowing the algorithm to extend to previously
unseen documents.

We are first interested in computing topic distributions for each
document in our corpus. The inference process of LDA estimates a
matrix $\theta$ where each row $\theta_d$ represents a mixture
distribution over $K$ latent topics for each document $d$. Given this
matrix, for each subreddit $S$, we can find the average topic
distribution of that subreddit as $\bar\theta_S = \frac{1}{|S|}
\sum_{d \in S} \theta_d$. In this case, we apply a topic model in the
traditional sense, treating individual text submissions as documents,
and words as the discrete observations.

Given $\bar{\theta}_S$ for each subreddit $S$, we define the textual
similarity of of communities $A$ and $B$ in terms of the
Jensen-Shannon divergence. Specifically, our symmetric similarity
function is given as
\begin{equation}
\label{eq:textsim}
S_{text}(A, B) = 1 - \frac{1}{2} \left(KL(\bar{\theta}_A || M)+KL(\bar
{\theta}_B || M)\right) \, ,
\end{equation}
where $M = \frac{1}{2} (\bar{\theta}_A + \bar{\theta}_B)$, and
$KL(X||Y)$ is the Kullback-Leibler divergence of $Y$ from $X$. Note
that $0 \leq S_{text}(A, B) \leq 1$.

We used the Mallet toolkit \cite{mccallum2002mallet} to perform
inference. We used a uniform Dirichlet prior over the topic-word
distributions of $\beta = .01$, and use the built-in functionality
for hyperparameter optimization over the document-topic prior $\alpha$
\cite{wallach2009rethinking}. We choose our number of topics $K$ by
sweeping the parameter value over a small set of values, namely,
$\{100, 300, 500\}$. Evaluating the quality of topic models is a
difficult task. For instance, it is known that topic models that fit
to unseen data better likely produce \emph{worse} topics, as judged by
human evaluators \cite{chang2009reading}. Here, we perform no
intrinsic evaluation of our models, deferring to our task-specific
parameter search with ground truth data to determine which number of
topics is best. A random sample of topics from the $K=300$ model is
given in Table \ref{tab:topics}.
\begin{table}[t]
\centering
\begin{tabular}{lll}
\toprule
$\alpha_k \cdot 10^2$ & Description & Top Words\\
\midrule
.992 & Purchase & store buy online stores shop find local sell good price\\
.433 & Donations & donate money charity raise donations people support donation\\
.332 & Tabletop RPG & character magic party level campaign spell spells dragon\\
.147 & Bioshock & time timeline booker elizabeth peter universe infinite end back\\
.133 & Pokemon & shiny male adamant timid female ball modest egg jolly traded\\
\bottomrule
\end{tabular}
\caption{A random sample of 5 topics from our LDA model learned from
  6.6M posts to the site \texttt{reddit.com}, along with
  human-authored descriptions. Also included are the optimized
  document-topic priors for each topic; this can be thought of as a
  rough indication of how frequently the topic appeared throughout the
  documents.}
\label{tab:topics}
\end{table}
\subsection{User Similarity}

While clustering
could be used to define user
similarity between different communities, we err on the side of
simplicity and use a
member-overlap-based
comparison as our starting
point. Specifically, we define the weight between communities $A$ and
$B$ in terms of their user sets $A_u$ and $B_u$ as the Jaccard
similarity given by
\begin{equation}
\label{eq:jac}
S_{user}(A,B) = \frac{|A_u \cap B_u|}{|A_u \cup B_u|} \, .
\end{equation}

\section{Network Clustering with Ground Truth}

Our first goal is to establish that these similarity metrics are able
to define networks that express community structure among subreddits
close to a ground-truth set of relationships we expect. We use an
off-the-shelf algorithm to cluster
based on
both the
text and user similarities
described in the previous section, and
compare against a set of hand-curated ground truth clusters. Once we
establish that these two networks express meaningful relationships, we
then discuss our method for latent interest detection.

We first gather a set of ground truth clusters of subreddits. Each
subreddit is associated with meta-information compiled by moderators
of that subreddit. Often included in this meta-information is a list
of related communities,\footnote{
An interactive visualization of inter-community connections based on this
related-subreddit information has been created by Andrei Kashcha, e.g.,
{\small\url{http://www.yasiv.com/reddit\#/Search?q=sanjosesharks}}.
}
 and we extracted 51 such connected components from these
lists, using several popular subreddits as starting points.
 After
filtering these lists for communities that were among the 3.2K we
considered, we were left with 37 ground truth clusters.

Standard, non-network clustering algorithms are not sufficient to
address our setting because of the \emph{overlapping} community
phenomenon. Traditional community detection algorithms
(\cite{lancichinetti2009community,leskovec2010empirical} offer good
reviews) generally assume that each node is a member of a
\emph{single} community. However, there is growing interest in
relaxing this assumption and allowing for cluster overlaps in the case
of complex
networks
\cite{kelley2009existence,xie2013overlapping,reid2013partitioning}.

In our case, it is very easy to think of cases where one subreddit
could reasonably belong to multiple clusters. For instance, consider
the subreddit \texttt{/r/SanJoseSharks}, which is dedicated to a
professional hockey franchise based in San Jose, California.
Allowing an
unsupervised algorithm the option to place this community into two
clusters, one for hockey teams and one for all California sports
teams, is reasonable. As such, we use a state-of-the-art overlapping
community
detector, Speaker-listener Label Propagation Algorithm (SLPA, also called
GANXiS) \cite {xie2012towards}, for our clustering.

SLPA outputs a set of overlapping clusters that we would like to compare
with a ground truth set of overlapping clusters. However,
typical
clustering evaluation metrics do not work if the single-membership
assumption is violated. For evaluation, we
therefore
use two overlapping-community evaluation metrics:
an extension of
\emph{normalized mutual information (NMI)} that accounts for
multi-cluster membership \cite{lancichinetti2009community} and the
\emph{Omega index $\Omega$} \cite{collins1988omega}.
Their
implementation is described and provided by \cite{McDaidNMI}.

\subsection{Re-scaling Similarities}

While a majority of pairwise
user-based similarities are zero, the median
text-based similarity between all pairs of subreddits computed from Equation
\ref{eq:textsim} is still very large. If a graph were constructed
using these unscaled, raw values, subreddit pairs with \emph{below
  average} textual similarity would still be assigned a positive
weight.

We compute the following rescaling of text similarities that is more
appropriately considered as a weight in a (sparser) graph. First, we
compute the mean $\mu$ of all $S_{text}(A,B)$. Then, if $S_{text}(A,B)
< \mu$, meaning that $A$ and $B$ have below average textual
similarity, the corresponding weight in the content network between
$A$ and $B$ is set to zero. If $S_{text}(A,B) \geq \mu$, $\mu$ is
subtracted from $S_{text}(A,B)$. Finally, the result is linearly
scaled such that $\mu$ maps to 0, and the maximum possible value maps
to $1$. In total, this sparsity-inducing re-scaling can be summarized
as:
\begin{equation}
\label{eq:rescale1}
S'_{text}(A, B) = \max\left(0,\, \frac{S_{text}(A, B)-\mu}{1-\mu}\right).
\end{equation}

Even after rescaling the text in accordance with Equation
\ref{eq:rescale1}, it is not clear that $S'_{text}$ and $S_{user}$
are, in their unmodified form, optimal for deriving network
weights. We introduce some scaling parameters which we optimize using
a validation set. Specifically, we partition our 37 ground truth
subreddit clusters into a validation set of 17 and a test set of
20. We perform a grid search over a percentile-cutoff parameter $p$
(i.e. edges are disregarded if they are under a specific percentile),
an exponential scaling factor $a$, a community overlapping propensity
measure $r$, and, in the case of the content graph, over the number of
topics $K$ included in the topic model. Edge weights that exceed the
percentile cutoff are then set according to the scaling factor as
\begin{equation}
\label{eq:finalWeights}
w(A,B) = exp(a \cdot S(A,B)) - 1 \,,
\end{equation}
where $S$ is $S'_{text}$ or $S_{user}$, depending on the context. $r$
is a parameter internal to SLPA. Because our validation/testing sets
are small, we run our experiments over 100 val/test splits.

\subsection{Experimental Results}

Table \ref{tab:results} compares methods of deriving network weights
against two baselines.  ``Random Const Size'' simply predicts a random
set of constant size clusters. ``Random True Size'' is allowed to
observe the size of the ground-truth sets, and generates a random
permutation preserving those sizes. The results reported are 95\%
confidence intervals computed over the 100 cross-validation splits.

Our text-based similarities perform better than the random baselines
and the user network. This result demonstrates that a topic model can
successfully be used to define a textual similarity function between
two complex communities, though simpler language comparison methods
might suffice. The user similarity network underperforms relative to
the content network, but this result is not entirely surprising. The
underlying ground truth was based on annotations provided by
moderators from particular communities reporting other communities
with similar \emph{content}. It is precisely these differences we wish
to extract with latent interest detection.

\begin{table}
\centering
\begin{tabular}{lcccccc}
\toprule
 & Random Const Size & Random True Size & Users & Text \\
\midrule
$100\cdot \Omega$ & $.74 \rpm .08$ & $.89 \rpm .10$ & $15.90 \rpm .57$ &
\textbf{51.36} $\rpm 2.40$ \\
\midrule
$100\cdot NMI$ & $0.0 \rpm 0.01$ & $0.0 \rpm 0.01$ & $18.51 \rpm .61$ &
\textbf{28.90} \rpm 1.50 \\
\bottomrule
\end{tabular}
\caption{Clustering evaluation results for two baselines, the user
  network, and the textual content network.
``Random Const Size'' is a
  constant prediction of constant-sized, randomly selected clusters. The constant size
  is selected to be the average number of subreddits that appear in a ground truth cluster,
  but membership is randomized.
``Random True Size'' first samples a random permutation of
the evaluation subreddits, and then partitions the permutation into
sets of equal to the size to the ground truth sets. ``Users'' is
community detection derived from pairwise Jaccard similarity scores
between user sets. ``Text'' is a content-based clustering derived from
textual similarity. All results are reported with 95\% confidence
intervals drawn over 100 random test splits. The maximum value for
both evaluation metrics is 100, higher is better.}
\label{tab:results}
\end{table}

\section{Latent Interest Detection}
\begin{table}[t]
\centering
\begin{tabular}{lp{2cm}p{8cm}}
\toprule
Community & Top Topics & Top Latent Interests \\\midrule
vegan & diet food cooking animal moral & AnimalRights Anarchism yoga VegRecipes Feminism environment \textbf{philosophy} \textbf{gardening} \textbf{bicycling} Buddhism \\\midrule \midrule

Liberal & elections government1 government2 arguments gunlaws & \textbf{California} ${\rm  GunsAreCool}^*$ \textbf{Bad\_Cop\_No\_Donut} economy \textbf{Feminism} \textbf{immigration} \textbf{RenewableEnergy} \textbf{energy} \textbf{newyork} democrats \\
 & & [${ }^*${\em ``GunsAreCool'' is a satirical name; this subreddit advocates stricter gun conrol}] \\
\midrule

Conservative & elections government2 government1 arguments legislation & \textbf{Bad\_Cop\_No\_Donut} \textbf{guns} \textbf{Christianity} \textbf{Military} \textbf{economy} Economics \textbf{Catholicism} progun \textbf{climateskeptics} \textbf{religion}\\
\midrule \midrule

SanJoseSharks & game hockey tickets win season & SFGiants SanJose 49ers \textbf{bayarea} OaklandAthletics \textbf{warriors} \textbf{EA\_NHL} \textbf{hockeyplayers} \textbf{SJSU} \textbf{SFBayJobs} \\

\midrule
CanadaPolitics & elections gunlaws discussion economy reddit & Quebec ontario metacanada toronto ottawa \textbf{Habs} montreal vancouver VictoriaBC \textbf{PersonalFinanceCanada}\\

\midrule
LadiesofScience & gradschool jobs college research socialLife  & labrats xxfitness femalefashionadvice \textbf{London\_homes} \textbf{askgis} bioinformatics \textbf{FancyFollicles} \textbf{craftit} \textbf{chemhelp} \textbf{GirlGamers}\\ 

\midrule
PAX & tickets event fishing vacationSugguestions junk1 & PaxPassExchange SeaJobs LoLCodeTrade \textbf{bostonhousing} boardgames \textbf{gamesell} Seattle \textbf{LeagueOfGiving} \textbf{DnD} \textbf{gameswap}\\
\bottomrule
\end{tabular}
\caption{Latent interest examples. The second column gives
manually-assigned
names for the most frequent topics in a particular
  community. The third column gives the top 10 latent interests
(subreddit names).  The first row addresses the title's question, the
  next two rows are exploratory political examples, and the bottom 4
  rows are cases where multi-community membership is more easily
  discovered. There are often significant differences between the
  topics discussed in a community and the focus of their latent
  interests.  Bolded are latent interests not found in the top 10 most
  similar subreddits by only user similarity.
  }
\label{tab:latentInterest}
\end{table}

To detect the latent interests of a given subreddit, we identify
communities with high user similarity, but low textual similarity. For
this task, we return to our consideration of text/user similarity
given in Equations \ref{eq:textsim} and \ref{eq:jac},
respectively. The task of combining these measures is complicated by
the fact that their corresponding distributions have very different
shapes.

Here, we only aim to pose the problem of how to detect latent
interests and to offer preliminary, baseline results; we leave a
comparison of methods for latent interest detection to future work.
  As a simple starting point, we
first compute the top 100 most similar subreddits in terms of
userbases.  From this set, we discard any subreddit that is among the
top 500 most similar in terms of textual similarity. We are left with
a set of subreddits with highly similar users, but relatively distinct
language. Of these, we compute a ranking with a simple heuristic that
rewards differences between user and text similarities.
  Specifically, we define the latent interest of communities
$A$ and $B$ as
\begin{equation}
\label{eq:latentInterest}
LI(A,B) = S_{user}(A,B) \cdot \left(1-S_{text}(A, B)\right)
\end{equation}
where $S_{text}(A, B)$ is given in Equation \ref{eq:textsim} and
$S_{user}(A,B)$ is Jaccard similarity as in Equation \ref{eq:jac}. The
simplicity of this formula is meant to convey optimization of a
mathematical conjunction. We maximize both similarity of user bases
and dissimilarity of text content without permitting one of these
objectives to overpower the other, unlike an additive formula of the
form $S_{user}(A, B) - S_{text}(A, B)$.

It should be noted that for subreddits with multiple plausible
memberships (i.e. \texttt{SanJoseSharks} could be considered in the
frame of ice hockey, or California sports) it is not meaningful to
declare one membership as the latent one apriori. To address this
ambiguity, we report the top topics from $\bar\theta_S$ along with the
detected latent interest. Ideally, it should be clear what the primary
topics of conversation are based on the topics discussed in the
text. We picked a set of 4 communities with clear multi-community
memberships to examine as a baseline. The latent interests derived in
these cases should be
clear,
yet should still contrast with
the main topical focus of the subreddit. These baselines are presented
in the last four rows of Table \ref{tab:latentInterest}. Because the
goal of latent interest detection is to discover unexpected and
surprising relationships, quantitative evaluation is a difficult
problem we leave to future work. It should be noted that we tuned our
ranking method by examining the model's output on the
\texttt{/r/Conservative}
community, and \texttt{/r/vegan} was chosen post-hoc; however, we
committed to our other specific example communities before seeing the
the results, and computed latent interests exactly once.  

Our method of contrasting textual and user similarity produces results
that are different than simply using a single ranking metric. For
instance, consider
the top latent interests of \texttt{/r/Liberal},
shown in Table \ref{tab:latentInterest}.
Of
those presented, 7/10 do not appear in the top 10 text/user similarity
rankings. In the case
of \texttt{/r/Conservative}, this fraction of novel discoveries is
8/10. By explicitly seeking subreddits with dissimilar content but
similar users, we discover new types of relationships.

\section{Conclusion}

We define two different similarity functions over networks with nodes
consisting of entire communities. We then use these definitions for a
graph clustering task to demonstrate their informativeness for latent
interest detection. We experimentally determine that anomalous
community relationships have face validity, but defer a more rigorous
quantitative evaluation to future work.

This work advances our ability to study social network behavior at
both the macro and micro scales. As the size and complexity of social
networks increases, understanding not just user-user interactions but
community-community interactions becomes increasingly important in
recognizing large-scale patterns. We can also use community-community
interactions to study small-scale behaviors at the user level, as
individuals select distinct forums to participate in distinct themes
and social roles --- even though the actual user community might be
nearly identical.

\subsubsection*{Acknowledgments}

This material is based upon work supported by the National Science
Foundation under Grant No. 0910664, a Google Research grant, the
Cornell Institute for the Social Sciences, and a Cornell University
fellowship. Any opinions, findings, and conclusions or recommendations
expressed in this material are those of the author(s) and do not
necessarily reflect the views of the supporting institutions.

\bibliography{refs}

\end{document}